\documentstyle[12pt]{article}
\newcommand{\nc}{\newcommand}
\nc{\noi}{\noindent}
\nc{\Jbf}{\mbox{\boldmath $J$}}
\nc{\erm}{{\rm e}}  \nc{\Rcal}{{\cal R}}    \nc{\rrm}{{\rm r}}
\nc{\AR}{A_{\rm 1R}} \nc{\ARAR}{A_{\rm 2R}}
\nc{\AT}{A_{\rm 1T}} \nc{\ATAT}{A_{\rm 2T}}
\nc{\al}{\alpha_1} \nc{\alal}{\alpha_2}      \nc{\Ga}{\Gamma}
\nc{\be}{\beta_1} \nc{\bebe}{\beta_2}
\nc{\psiuno}{\psi_{\rm I}} \nc{\psidue}{\psi_{\rm II}}
\nc{\psitre}{\psi_{\rm III}} \nc{\psiquattro}{\psi_{\rm IV}}
\nc{\psicinque}{\psi_{\rm V}}
\nc{\bb}{\begin{equation}} \nc{\ee}{\end{equation}}
\nc{\um}{{1\over 2}} \nc{\C}{I\!\!\!C} \nc{\R}{I\!\!R}
\nc{\pa}{\partial} \nc{\ug}{\; = \;}
\nc{\cent}{\centerline} \nc{\vs}{\vspace*}

\begin{document}

\baselineskip 0.7cm

\cent{\bf Superluminal tunneling through two successive barriers}
\footnotetext{$^{(\dagger)}$ Work partially supported by MURST, INFN and CNR
(Italy), by CAPES (Brazil), and by I.N.R. (Kiev) and S.K.S.T. of
Ukraine. \ [Appeared in preliminary version as Lanl Archives
\# quant-ph/0002022].}

\begin{center}
{Vladislav S. Olkhovsky\break
{\em Institute for Nuclear Research, Kiev-03028;
Research Center "Vidhuk", Kiev, Ukraine;\break
e-mail: {\rm olkhovsk@stenos.kiev.ua}}}\\

\

{Erasmo Recami\break
{\em Facolt\`a di Ingegneria, Universit\`a statale di Bergamo,
24044--Dalmine (BG), Italy;\break
I.N.F.N.--Sezione di Milano, Milan, Italy; \ {\rm and}\break
C.C.S., State University at Campinas, Campinas, S.P., Brazil;\break
e-mail: {\rm recami@mi.infn.it}}}\\

and\\

{Giovanni Salesi\break
{\em I.N.F.N.--Sezione di Milano, Milan, Italy; \ {\rm and}\break
Facolt\`a di Ingegneria, Universit\`a statale di Bergamo, 
24044--Dalmine (BG), Italy;\break
e-mail: {\rm salesi@ct.infn.it}}}
\end{center}

\noi {\bf Abstract ---} We study the phenomenon of one-dimensional
non-resonant tunneling through two successive (opaque) potential barriers, 
separated by an intermediate free region $\Rcal$, by analyzing the relevant 
solutions to the Schroedinger equation. \ We find that the total traversal 
time does {\em not} depend not only on the barrier widths (the so-called 
``Hartman effect"), but also on the $\Rcal$ width: so that the effective 
velocity in the region $\Rcal$, between the two barriers, can be regarded 
as practically infinite. \ This agrees with the results known from the
corresponding waveguide experiments, which simulated the tunneling experiment
herein considered due to the known formal identity between the Schroedinger
and the Helmholtz equation.\\

\noi PACS numbers: \ 73.40.Gk ; \ 03.65.-w ; \ 03.30.+p ; \ 41.20.Jb ;
\ 84.40.Az

\newpage

{\bf 1. -- Introduction ---} \ It is known within quantum mechanics, with
regard to the tunneling processes, that the tunneling time ---either evaluated
as a simple ``phase time''[1] or calculated through the analysis of the
wavepacket behaviour[2]--- does not depend on the barrier width in the case of
opaque barriers.  Such a phenomenon, sometimes called ``Hartman effect''[3],
implies Superluminal and arbitrarily large (group) velocities $v$ inside
long enough barriers[2]. Experiments that may verify this prediction by,
say, electrons are difficult.  Luckily enough, however, the Schroedinger
equation in the presence of a potential barrier is mathematically
identical[4] to the Helmholtz equation for an electromagnetic wave
propagating, e.g., down a metallic
waveguide along the $x$-axis: and a barrier height $V$ bigger than the
electron energy $E$ corresponds (for a given wave frequency) to a waveguide
transverse size smaller than a cut-off value. A segment of {\em undersized}
guide does therefore behave as a barrier for the wave (photonic barrier): So
that the wave assumes therein ---like an electron inside a quantum
barrier--- an imaginary momentum or wave-number and gets exponentially damped
along $x$, as a consequence. In other words, it becomes an {\em evanescent}
wave (going back to normal propagation, even if with reduced amplitude,
when the narrowing ends and the guide returns to its initial transverse
size). \ Thus, a tunneling experiment can be simulated by having
recourse to evanescent waves (for which the concept of group velocity can be
properly extended[5]).

And the fact that evanescent waves travel with
Superluminal speeds has been actually {\em verified} in a series of famous
experiments. Namely, various experiments ---performed since 1992 onwards
by R.Chiao's and A.Steinberg's group at Berkeley[6], by G.Nimtz at
Cologne[7], by A.Ranfagni and colleagues at Florence[7], and by others at
Vienna, Orsay, Rennes[7]--- verified that ``tunneling photons" travel with
Superluminal group velocities; in other words, they confirmed, directly or
indirectly, the occurrence of the Hartman effect.

Let us emphasize that the most interesting experimental setup, dealing 
with evanescent waves, seems to be ---however--- the one comprehending 
{\em two} successive {\em evanescence regions} (``classical barriers"),
separated by a segment of normal region.  For suitable frequency bands
---i.e., far from resonances---, it was found that the total crossing time
does not depend on the length of the intermediate (normal) region: namely,
that the beam speed along it is infinite.  The related experimental
results[8] have been already confirmed by numerical simulations, based on
Maxwell equations only[9]. \ But they are so amazing that we want to check
whether they agree also with what is predicted by quantum mechanics in
the analogous case of two successive potential barriers.

In this note we are actually going to show that, for non-resonant tunneling
trough two successive, rectangular (opaque) potential barriers (Fig.1), 
the (total) phase time does depend neither on the barrier widths
{\em nor on the distance between the barriers}. In other words, far from
resonances the tunneling phase time, which does depend on the entering
energy, can be shown to be {\em independent} of the distance between the two
barriers.

\

{\bf 2. -- Phase time evaluation ---} \ Let us consider the (quantum-mechanical)
stationary solution for the one-dimensional (1D) tunneling of a
non-relativistic particle, with mass $m$
and kinetic energy \ $E=\hbar^2k^2/2m=mv^2/2$, \ through two equal
rectangular barriers with height $V_0$ ($V_0>E$) and width $a$, \ quantity
$L-a \ge 0$ being the distance between them. \ The Schr\"odinger equation is

$$
-\frac{\hbar^2}{2m}\,\frac{\pa^2}{\pa x^2}\,\psi(x) \; + \; V(x)\,\psi(x)
\ug E\,\psi(x)\,,
\eqno{(1)}
$$

\noi where $V(x)$ is zero outside the barriers, while $V(x) = V_0$ inside the
potential barriers. In the various regions I~$(x\leq 0)$, II~$(0\leq x\leq a)$,
III~$(a\leq x\leq L)$, IV~$(L\leq x\leq L+a)$ and V $(x\geq L+a)$, the
stationary solutions to eq.(1) are the following

\

$\hfill{\displaystyle\left\{\begin{array}{l}
\psiuno \ug \erm^{+ikx} + \AR\,\erm^{-ikx}\\

\psidue \ug \al\,\erm^{-\chi x} + \be\,\erm^{+\chi x}\\

\psitre \ug \AT\,\left[\erm^{ikx} + \ARAR\,\erm^{-ikx}\right]\\

\psiquattro \ug \AT\,\left[\alal\,\erm^{-\chi(x-L)} +
\bebe\,\erm^{+\chi(x-L)}\right]\\

\psicinque \ug \AT\,\ATAT\,\erm^{ikx} \; ,
\end{array}\right.}
\hfill{\displaystyle\begin{array}{r}
(2{\rm a}) \\ (2{\rm b}) \\ (2{\rm c}) \\ (2{\rm d}) \\ (2{\rm e})\end{array}}$

\

\

\noi where $\chi\equiv\sqrt{2m(V_0-E)}/\hbar$, and quantities $\AR$, $\ARAR$,
$\AT$, $\ATAT$, $\al$, $\alal$, $\be$ and $\bebe$ are the reflection
amplitudes, the transmission amplitudes, and the coefficients of the
``evanescent" (decreasing) and ``anti-evanescent" (increasing) waves for
barriers 1 and 2, respectively. \ Such quantities can be easily obtained from
the matching (continuity) conditions:

\

$\hfill{\displaystyle\left\{\begin{array}{l}
\psiuno(0) \ug \psidue(0)\\
{\displaystyle\left.\frac{\pa\psiuno}{\pa x}\right|_{x=0}} \ug
{\displaystyle\left.\frac{\pa\psidue}{\pa x}\right|_{x=0}}
\end{array}\right.}
\hfill{\displaystyle\begin{array}{r}
(3{\rm a}) \\ (3{\rm b})\end{array}}$

\

\

$\hfill{\displaystyle\left\{\begin{array}{l}
\psidue(a) \ug \psitre(a)\\
{\displaystyle\left.\frac{\pa\psidue}{\pa x}\right|_{x=a}} \ug
{\displaystyle\left.\frac{\pa\psitre}{\pa x}\right|_{x=a}}
\end{array}\right.}
\hfill{\displaystyle\begin{array}{r}
(4{\rm a}) \\ (4{\rm b})\end{array}}$

\

\

$\hfill{\displaystyle\left\{\begin{array}{l}
\psitre(L) \ug \psiquattro(L)\\
{\displaystyle\left.\frac{\pa\psitre}{\pa x}\right|_{x=L}} \ug
{\displaystyle\left.\frac{\pa\psiquattro}{\pa x}\right|_{x=L}}
\end{array}\right.}
\hfill{\displaystyle\begin{array}{r}
(5{\rm a}) \\ (5{\rm b})\end{array}}$

\

\

$\hfill{\displaystyle\left\{\begin{array}{l}
\psiquattro(L+a) \ug \psicinque(L+a)\\
{\displaystyle\left.\frac{\pa\psiquattro}{\pa x}\right|_{x=L+a}} \ug
{\displaystyle\left.\frac{\pa\psicinque}{\pa x}\right|_{x=L+a}}
\end{array}\right.}
\hfill{\displaystyle\begin{array}{r}
(6{\rm a}) \\ (6{\rm b})\end{array}}$

\

\

\begin{figure}
\begin{center}
\setlength{\unitlength}{0.240900pt}
\ifx\plotpoint\undefined\newsavebox{\plotpoint}\fi
\sbox{\plotpoint}{\rule[-0.200pt]{0.400pt}{0.400pt}}%
\begin{picture}(1500,900)(0,0)
\font\gnuplot=cmr10 at 10pt
\gnuplot
\sbox{\plotpoint}{\rule[-0.200pt]{0.400pt}{0.400pt}}%
\put(281,469){\makebox(0,0){I}}
\put(491,469){\makebox(0,0){II}}
\put(806,469){\makebox(0,0){III}}
\put(1121,469){\makebox(0,0){IV}}
\put(1331,469){\makebox(0,0){V}}
\put(428,574){\makebox(0,0)[l]{1}}
\put(1058,574){\makebox(0,0)[l]{2}}
\put(323,603){\makebox(0,0)[l]{$V_0$}}
\put(953,603){\makebox(0,0)[l]{$V_0$}}
\put(386,431){\makebox(0,0){$0$}}
\put(596,431){\makebox(0,0){$a$}}
\put(1016,431){\makebox(0,0){$L$}}
\put(1226,431){\makebox(0,0){$L+a$}}
\put(176,450){\usebox{\plotpoint}}
\put(385.17,450){\rule{0.400pt}{30.100pt}}
\multiput(384.17,450.00)(2.000,87.526){2}{\rule{0.400pt}{15.050pt}}
\put(176.0,450.0){\rule[-0.200pt]{50.348pt}{0.400pt}}
\put(594.67,450){\rule{0.400pt}{36.135pt}}
\multiput(594.17,525.00)(1.000,-75.000){2}{\rule{0.400pt}{18.067pt}}
\put(387.0,600.0){\rule[-0.200pt]{50.107pt}{0.400pt}}
\put(1015.67,450){\rule{0.400pt}{36.135pt}}
\multiput(1015.17,450.00)(1.000,75.000){2}{\rule{0.400pt}{18.067pt}}
\put(596.0,450.0){\rule[-0.200pt]{101.178pt}{0.400pt}}
\put(1225.17,450){\rule{0.400pt}{30.100pt}}
\multiput(1224.17,537.53)(2.000,-87.526){2}{\rule{0.400pt}{15.050pt}}
\put(1017.0,600.0){\rule[-0.200pt]{50.107pt}{0.400pt}}
\put(1227.0,450.0){\rule[-0.200pt]{50.348pt}{0.400pt}}
\end{picture}
\end{center}
\caption{The non-resonant tunneling process, through {\em two} successive
(opaque) potential
barriers, considered in this paper. We show that, far from resonances,
the (total) phase time for tunneling through the two barriers does depend
neither on the barrier widths {\em nor on the distance between the
barriers}.}
\end{figure}
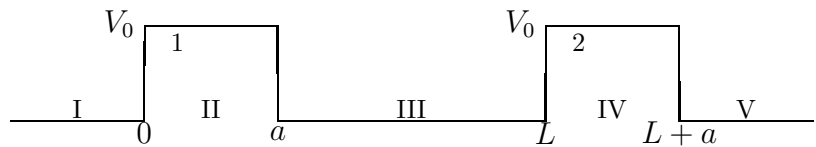

\

Equations (3-6) are eight equations for our eight unknowns ($\AR$, $\ARAR$,
$\AT$, $\ATAT$, $\al$, $\alal$, $\be$ and $\bebe$). \ First, let us obtain
the four unknowns $\ARAR$, $\ATAT$, $\alal$, $\bebe$ from eqs.(5) and (6)
in the case of {\em opaque} barriers, i.e., when $a$ is large enough (and $\chi$
not too small) so that one can assume that $\chi a\rightarrow\infty$:

\

$\hfill{\displaystyle\left\{\begin{array}{lcr}
\alal \longrightarrow \erm^{ikL}\,\displaystyle{\frac{2ik}{ik-\chi}}& \
\hskip 4cm & (7{\rm a})\\

\bebe \longrightarrow \erm^{ikL-2 \chi a}\,\displaystyle{\frac{-2ik(ik+\chi)}
{(ik-\chi)^2}}& \ & (7{\rm b})\\

\ARAR \longrightarrow \erm^{2ikL}\,\displaystyle{\frac{ik+\chi}{ik-\chi}}& \
& (7{\rm c})\\

\ATAT \longrightarrow \erm^{-\chi a}\erm^{-ika}\,\displaystyle
{\frac{-4ik\chi}{(ik-\chi)^2}} & \ & (7{\rm d})
\end{array}\right.}$

\

Then, we may obtain the other four unknowns $\AR$, $\AT$, $\al$, $\be$ from
eqs.(3) and (4). Aagain in the case of large enough barriers
(and $\chi a\rightarrow\infty$), one gets:

\

$\hfill{\displaystyle\left\{\begin{array}{lcr}
\al \longrightarrow \displaystyle{\frac{2ik}{ik-\chi}}& \
\hskip 4cm & (8{\rm a})\\

\be \longrightarrow \erm^{-2\chi a}(k-i\chi)
\displaystyle{\frac{\sin k(L-a)}{\chi}}A& \
& (8{\rm b})\\

\AR \longrightarrow \displaystyle{\frac{ik+\chi}{ik-\chi}}& \
& (8{\rm c})\\

\AT \longrightarrow \erm^{-\chi a}\erm^{-ikL}A\,,& \
& (8{\rm d})
\end{array}\right.}$

\

\noi where

$$
A \equiv \displaystyle{{2 \chi k} \over {2 \chi k \; \cos k(L-a) \, + \,
(\chi^2 - k^2) \; \sin k(L-a)}} \,
\eqno{(9)}
$$

\noi results, incidentally, to be real.

At this point, by applying the well-known definition of phase-time
(see, for instance, refs.[1-3]), we can derive that the tunneling time

$$
{\tau^{\rm ph}}_{\rm tun} \equiv
\hbar\,\frac{\pa\arg\left[\AT\ATAT\erm^{ik(L+a)}\right]}{\pa E} \; =  \;
\hbar\,{\pa\over{\pa E}}\arg{\left[{{-4ik\chi}\over{(ik-\chi)^2}}\right]} =
$$

$$
=\hbar\,{\pa\over{\pa E}}\,\arctan\left[\frac{k^2-\chi^2}{k \: \chi}\right]
= \frac{1}{\hbar \chi} \; \frac{2m}{k} \, ,
\eqno{(10)}
$$

\noi while depending on the energy of the tunneling particle, {\em does not
depend} on $L+a$ (it being actually independent both of $a$ and of $L$).

This result does {\em not only} confirm the so-called ``Hartman effect"[2,3]
for the two opaque barriers ---i.e., the independence of the tunneling
time from the opaque barrier widths,--- but it does {\em also} extend 
such an effect by implying the total tunneling time to be independent even 
of $L$ (see Fig.1).  This might be regarded as a further evidence of
the fact that quantum systems seem to behave as non-local; but is has a
more general meaning, it being associated with the properties of any waves
(and, in fact, something very similar happens also, e.g., with electromagnetic
waves: see below). \ It is important to stress once more that the previous
result holds, however, for non-resonant (nr) tunneling: \ i.e., for energies
far from the resonances that arise in region
III due to interference between forward and backward traveling waves (a
phenomemon quite analogous to the Fabry-P\'erot one in the case of classical
waves). Otherwise it is known that the expression for the time delay
$\tau$ near a resonance is rather larger: for example, for a
gaussian resonance at $E_\rrm$ with half-width $\Ga$, it would be \
$\tau \: = \: \hbar \Ga [(E - E_\rrm)^2 + \Ga^2]^{-1} \: + \: \tau_{\rm nr}$. 

{\bf 3. -- Discussion ---} \ The tunneling-time independence from the width
($a$) of each one of the two
opaque barriers is itself a generalization of the Hartman effect, and might be 
a priori understood ---following refs.[4,6]--- on the basis of the reshaping
phenomenon which takes place inside a barrier.

With regard to the even more interesting tunneling-time independence from the
distance $L-a$ between the two barriers, it may be understood on the basis of
the interference {\em between the waves} coming out of the first barrier
(region II) and traveling in region III {\em and the waves} reflected from
the second barrier (region IV) back into the same region III.

Such an interference has been shown[2] to cause an ``advancement", i.e.,
an effective acceleration of the forward-traveling waves, even in
region I: Namely, going on to
the wavepacket language, we noticed in refs.[2] that the arriving wavepacket
does interfere with the reflected waves that start to be generated as soon
as the packet forward tail reaches the first barrier edge: In such a way
that, already before the barrier, the backward tail of the initial wavepacket
decreases ---because of destructive interference with those reflected waves---
at a larger degree than the forward one. This simulates an increase of the
average speed of the entering packet; hence, the effective (average)
flight-time of the approaching packet from the source to the barrier does
decrease.

So, a reshaping and ``advancement" of the same kind (inside the barriers, as
well as to the left of the barriers) may qualitatively explain why the
tunneling-time is independent of the barrier widths and of the distance
between the two barriers.  Phenomena of this kind, actually, do not seem
to be at variance with Special Relativity, as it has already been discussed
in a number of papers (cf., e.g., refs.[9] and [5], and refs. therein). \
 \ It remains impressive, nevertheless, that
in region III ---where no potential barrier is present, the current  is
non-zero and the wavefunction is oscillatory,--- the effective speed
(or group-velocity) is practically {\em infinite}.\footnote{Loosely speaking,
one might say that the considerd two-barriers setup {\em can} behave as a
``space destroyer" with reference to its intermediate region.} 
After some straightforward but rather bulky calculations,
one can moreover see that the same effects (i.e., the independence from the 
barrier widths and from the distances between the barriers) are still valid 
for any number of barriers, with different widths and different distances 
between them.

Finally, let us recall that the known similarity between photon and
(nonrelativistic) particle tunneling[2,4,10,11] implies our previous results to
hold also for photon tunneling through successive ``barriers": For example,
for photons in presence of two successive band gap filters, like two suitable
gratings or two photonic crystals. Experiments should be easily realizable;
while indirect experimental evidence seems to come from papers such as [12].

Let us also repeat that the classical, relativistic (stationary) Helmholtz
equation for an electromagnetic wavepacket in a waveguide is known to be
formally identical to the quantum, non-relativistic (stationary) Schroedinger
equation for a potential barrier;\footnote{These
equations are however different (due to the different order
of the time derivative) in the time-dependent case. Nevertheless, it can be
shown that they still have in common classes of analogous solutions,
differing only in their spreading properties[2,11].} \ so that, for instance,
the tunneling of a particle under and along a barrier has been 
simulated[2,4,7-11,13] by the traveling of evanescent waves along an undersized
waveguide. Therefore, the results of this paper are to be valid also for
electromagnetic wave propagation along waveguides with a succession of
undersized segments (the ``barriers") and of normal-sized segments. This
confirms the results obtained, within the classical realm, directly
from Maxwell equations[9,13], as well as by the known series of
``tunneling" experiments performed ---till now--- with microwaves (see
refs.[7] and particularly [8]).

\

\

{\bf Acknowledgements}

\noi The authors are grateful to S.Esposito, as well as to A.Agresti,
G.G.N.Angilella, P.Bandyopadhyay, F.Bassani, G.Benedek, R.Bonifacio,
L.Bosi, C.Becchi, G.C.Cavalleri, R.Y.Chiao,
R.Colombi, G.C.Costa, G.Degli Antoni, P.Falsaperla,
F.Fontana, R.Garavaglia, A.Gigli Berzolari, H.E.Hern\'andez F., J.Jakiel,
E.Kapuschik, L.C.Kretly, G.Kurizki, G.Marchesini, R.Mignani, D.Mugnai,
G.Privitera, A.Ranfagni, A.Salanti, A.Shaarawi, A.Steinberg, J.W.Swart,
M.Zamboni-Rached and B.Zappa for scientific collaboration
or stimulating discussions; \ as well as to the Referees for very useful
remarks. \ At last, for the active cooperation special thanks are
due to G.Giuffrida, R.Salesi and M.T.Vasconselos. 

\

\

\end{document}